\documentstyle[12pt]{article}
\newcounter{mycount}

\newcommand{\be}{\begin{eqnarray}}
\newcommand{\ee}{\end{eqnarray}}

\newcommand\ie {{\it i.e. }}
\newcommand\eg {{\it e.g. }}

\newcommand\half{\frac 1 2 }

\newcommand\noi{\noindent}

\begin{document}

\centerline{\Large\bf Quantum mechanics on thin cylinders}
\vspace* {-25 mm}
\begin{flushright}  USITP-92-13 \\
December 1992
\end{flushright}
%\vskip 18mm
\vskip 30mm
\centerline{ Erik Westerberg and T. H. Hansson$^{\dagger}$ }
\vskip 4mm
\centerline {wes@vana.physto.se $ \ {\rm and} \ $ hansson@vana.physto.se }
\centerline{Institute of Theoretical Physics }
\centerline{University of Stockholm }
\centerline{ Vanadisv\"agen 9 }
\centerline { S-113 46 Stockholm}
\centerline{Sweden }
\vskip 2.0cm
\newcommand \sss {\mbox{ $<\overline{s}s>$} }
\def\fk{\mbox{ $f_K$} }
\centerline{\bf ABSTRACT}
\vskip 3mm

We discuss the quantum mechanics of particles of arbitrary statistics on an
infinite cylinder with and without a magnetic field perpendicular to the
surface. In the presence of a magnetic field, the
translational symmetry along the cylinder is broken down to a discrete one
by  the Aharonov-Bohm effect. For interacting fermions in a strong field we
get an  effective one-dimensional lattice model that in a limit can be
mapped on an  Ising chain. We also show that a system of anyons on a
cylinder are,  in a certain limit closely related to the 1-dimensional
(integrable)  Sutherland model. By
order of magnitude estimates we demonstrate that none of these effects are
likely to be experimentally observed with present techniques.

\vfil
\vskip 5mm
\noindent
$^{\dagger}$Supported by the Swedish Natural Science Research Council

\eject

\section{Introduction and summary}

It is now well known  that the quantum theory of identical particles in two
dimensions allows
for  fractional statistics \cite{lein1,wilc1,wilc6}. The statistics
is characterized by an angle $\theta$ that interpolates
continuously between bosons ($\theta = 0$) and fermions ($\theta
= \pi$). For particles moving on topologically non-trivial
surfaces, there are additional parameters that can be thought of
as magnetic fluxes through the non-contractible loops on the
surface. In this case the quantum mechanics is more complicated,
and in general there is an interplay between the phases due to
(fractional) statistics, and the Aharonov-Bohm  phases due to the
magnetic fluxes. In particular, for surfaces with handles there
are restrictions on the possible values of
$\theta$ \cite{eina1}. The simplest example of a non-simply
connected two-dimensional base space is the  cylinder, and the
aim of this paper is to analyze the quantum mechanics in this
case in some detail.

Since there has been some confusion concerning the possible  values of
$\theta$ on the
cylinder \cite{hats1}, section 2 contains a brief analysis of the
corresponding braid group.
We show that there are no restrictions on $\theta$,  and that there is a
single additional  arbitrary
angle, $\phi$, corresponding to the  magnetic flux  through the cylinder.
We then
construct the wave functions corresponding to any values of $\theta$ and
$\phi$, thus  explicitly showing that both $\theta$ or $\phi$ are arbitrary.

Next we consider particles with charge $q$ on a cylinder (of radius $R$)
with a homogenous magnetic field (of strength $B$) perpendicular to the
surface. This system looks translationally invariant, but we show that due
to the  Aharonov-Bohm effect  only the discrete symmetry $x\rightarrow x+n\,
\ell^2/R$ ($n$ an integer) is left, where $\ell = (qB)^{-\half}$ is the
magnetic length. This is easily  understood;  a net flux of $2\pi RB$ leaves
the cylinder through the surface per unit length, so the magnetic flux
inside the cylinder, which the particles moving on the surface feel via the
Aharonov-Bohm effect, decreases linearly along the axis. If this flux is
zero at some $x_0$, then adding an integer number $n$ of flux quanta $\Phi_0
=2\pi /q$ does not change the phases on the surface (and hence not the
physics), but shifts $x_0  \rightarrow x_0+ n\, \ell^2/R$ corresponding to
the discrete symmetry just mentioned.  %In this case the angle $\phi$ is
linearly related to $x_0$ and the %discrete symmetry $x \rightarrow n/qBR$
reflects the periodicity $\phi \rightarrow \phi +2\pi n$. The %remaining
symmetry is also manifest in the wave functions.

Could effects of breaking translational invariance be detected in
experiments?  Could we \eg imagine making a small cylinder in the lab, or
perhaps even using naturally occurring cylinders like buckytubes
\cite{iiji1}? These questions are addressed in section 5. Leaving aside the
(very hard) technical problem of how to bring the magnetic field out through
the surface of a cylinder,  (there is clearly no effect if this is not the
case) we  make order of  magnitude estimates of the parameters required to
detect any effect. The conclusions are negative; with present magnetic
fields we cannot hope even to insert sufficient amount of flux in a cylinder
thin enough to allow for the detection of broken translational symmetry.

Since our system is characterized by two energy scales, $1/mR^2$ and $qB/m$
(or length scales $R$ and $\ell$), it is in general quite
complicated and it is of obvious interest to study different limits where
one or both of these scales are large compared to other relevant scales like
the interparticle interaction  energy and thermal
excitation energies. In these situations it is natural to think in terms of
dimensional reduction, and describe the system by an effective
one-dimensional model.

The limit $R$ small and $R \ll \ell$  is
rather trivial for bosons or fermions; the particles are simply restricted
to move along the cylinder and the resulting theory is again bosons or
fermions  but on a line. For anyons the situation is a bit more complicated
since the corresponding one-dimensional ($1d$) model also exhibits
fractional statistics in a sense that was originally discussed in
\cite{lein1}. In section 4 we use the methods in  \cite{hans6} to derive
the  statistics parameter $\eta$ of the dimensionally reduced system.

The limit $\ell$ small  (\ie $qB$ large) and $\ell \ll R$  is
also rather trivial and simply corresponds to confining the particles
to the lowest Landau level. As for
the infinite plane we can write down the full anyonic lowest Landau level
spectrum and this is done in section 4.3.

The most interesting limit, which we shall study in some detail, is the
combination of the two above where $R$ and $\ell$ both are small.
In fact there will be a new relevant length scale $a=\ell^2/R$.
 We show that in this limit a system
of interacting fermions is described by a one  dimensional lattice model,
explicitly exhibiting the discrete
symmetry $x\rightarrow x+a$ discussed above.  We discuss some properties of
this lattice model and show that for any interaction, in
the "sharp" lattice limit  $qB  \rightarrow \infty $, $R  \rightarrow  0 $
and  $\ell^2/R  \rightarrow  a $, we
obtain a one-dimensional Ising chain.

For anyons moving in an infinite plane penetrated by a magnetic field, the
subset of the spectrum (wave functions and energies) corresponding to
anyonic continuation of  all fermi states in the lowest Landau level, is
exactly known \cite{halp1,karl1}. In the spirit of the  above, we expect
this subset to correspond to a one dimensional model. This is indeed the
case, and it has been shown \cite{hans6,polyc1,brin3} that anyons in the
lowest Landau level, with the conserved angular momentum acting as the
'Hamiltonian', is equivalent to the   Calogero model \ie to  particles on a
line interacting via the two body   potential $ \half m\omega^2(x_i-x_j)^2 +
g/(x_i-x_j)^2$ \cite{calo1,calo2}. In section 4.3 we ask whether a similar
construction is possible on the cylinder.  Indeed we find that in the
"sharp" lattice limit there is an equivalent integrable $1d$ model, similar
to the Calogero model, but defined on a circle and with the interparticle
interaction $g/\sin^2 (x_i-x_j)$. This model, first studied by Sutherland
\cite{suth3}, has the same eigenvalues and the same degeneracies as the
anyons on the cylinder.  We also make some comments about the  wave
functions, and conclude with a speculation concerning the corresponding
system on a torus.

\section{Quantum mechanics on a cylinder}
Following Leinaas and Myrheim \cite{lein1}, we quantize a
system of free particles by constructing wave functions that form a unitary
representation of the first homotopy group of
the classical configuration space. In the plane this leads to fractional
statistics  with an arbitrary phase $e^{i\theta}$ associated with
the interchange of the coordinates for two particles.
On topologically more complicated two dimensional surfaces there are
additional generators of the first homotopy group, corresponding to moving
particles around non-contractible loops  on the surface. Einarsson has shown
\cite{eina1,eina2}, that for closed surfaces this  imposes restrictions on
the statistical  phase and introduces new  phases associated with transport
around the handles of the surfaces.   He also showed that for anyons
($\theta \neq 0,\pi$) this requires the wave  function to have more than one
component.  In the case of the cylinder there is, in addition to the
particle exchange generators, a set of generators associated to paths which
take a particle once around the cylinder.
When quantizing there are two independent phases to be determined;
$e^{i\theta}$ determining the statistics and $e^{i\phi}$ which can be
interpreted  as an Aharonov-Bohm phase from a constant flux flowing through
the cylinder. There is no restriction on either of these phases \ie neither
on the statistics, nor on the flux.  The analysis of the braid group on the
cylinder is  well known but since there has been some confusion concerning
the resulting statistics we will briefly describe the construction. For a
general discussion we refer to\cite{birm1}.

 The classical configuration space of $N$ hard core identical particles on
the cylinder is

\be
Q^N & = & (C^N - \Delta )/P^N  \label{eq:QN}
\ee
where $C=S^1 \times R$ is
the cylinder, $\Delta = \{x\in C^N \mid x_i = x_j \  {\rm some} \ i\neq
j\}$ is the set of points of particle coincidence and $P^N$ the permutation
group of $N$ particles.
 The first homotopy group of $Q^N$ is generated by two classes of generators;
$\sigma_i$ which exchanges particle $i$ with particle $i+1$,
and $\rho_k$ which takes
particle $k$ once around the cylinder leaving particles $1..k-1$ to the
left and particles $k+1..N$ to the right  of the loop (Fig 1a). The
generators $\{\sigma_i,\rho_k\}$ obey the set of relations
\be
\sigma_k \sigma_l & = & \sigma_l \sigma_k \ \ \ \ \ \ \ \ \ \ \ \ \ \ |k-l|
\neq 1  \label{eq:generatorref1} \\
\sigma_k\sigma_{k+1}\sigma_k & = & \sigma_{k+1}\sigma_k\sigma_{k+1}
\label{eq:generatorref2} \\
\rho_i\rho_j & = & \rho_j\rho_i
\label{eq:generatorref3} \\
\rho_{i+1} & = & \sigma_i\rho_i\sigma_i
\label{eq:generatorref4}
\ee

Relation (\ref{eq:generatorref4}), displayed in Fig 1b, shows that we need
only one generator, say $\rho_1$, among the $\{\rho_k\}$.

Since higher dimensional representations of the generators correspond to
particles with some internal 'statistical' degree of freedom \cite{wilc6} we
restrict ourselves to one dimensional  representations of the braid group
(scalar anyons). Hence we represent the generators $\{\sigma_i,\rho_k\}$
with phases, and according to (\ref{eq:generatorref2}) the $\sigma_i$:s must
all be represented by the same 'statistical' phase which we customarily
denote by $e^{i\theta}$.  Writing $\rho_1 = e^{i\phi}$,
\footnote{
With a slight, but common, abuse of notation we use the same symbols for the
generators and their representations}
relation (\ref{eq:generatorref4}) immediately
implies
\be
\rho_k = \rho_1e^{i2(k-1)\theta} = e^{i[\phi + 2(k-1)\theta ]} \ \ .
\label{eq:rhokrel}
\ee
The remaining relations (\ref{eq:generatorref1}) and
(\ref{eq:generatorref3})  give no further
restriction on $\theta$ or $\phi$.
While the interpretation of $\theta$ as the statistics angle is clear, there
has been some confusion as to the meaning of relation  (\ref{eq:rhokrel}).
To clarify this point we recall that if we use single-valued wave functions,
the phases
representing the elements in the fundamental group  are  Wilson loops
\be
W[\Gamma ] = e^{i\oint_\Gamma dx A}
\label{eq:Wilsonloop}
\ee
where the vector potential $A$ is the connection that defines parallel
transport of position eigenstates on the Hilbert space of wave functions.
The phases $e^{i\theta}$ originate from a 'statistical' gauge potential,
while the phases associated with the $\rho_i$:s, where the integration is
around the cylinder, are related to the flux through  the cylinder. (In a
real experimental setup with charged particles, this is of course the real
magnetic flux.) It is now easy to understand  the $k$-dependence in
(\ref{eq:rhokrel}); assume that a flux $\phi$ leaves the left end of the
cylinder and that a  statistical flux $2\theta$ flows through each anyon
(Fig. 2).  The path which takes the $k^{th}$
particle around the cylinder will now enclose a total flux $\phi_{tot} =
\phi + (k-1)2\theta$, resulting in the phase factor (\ref{eq:rhokrel}).
\footnote{ In ref. \cite{hats1} the authors identify the generators
$\rho_1$ and $\rho_N^{-1}$, and erroneously conclude that there is a
constraint on the possible statistics
of particles moving on the cylinder. It should be clear from the analysis of
the braid group and the subsequent discussion that this
identification is incorrect. In (\ref{eq:cyltransform}) we also explicitly
construct wave functions  on the cylinder for any value of the statistical
parameter $\theta$.}

For anyons in a plane, characterized by an arbitrary
statistical angle $\theta$, there exist a singular gauge transformation
which maps a  system of bosons (with  single valued wave functions)
interacting via a $\theta$-dependent gauge potential, onto a system of
non-interacting (\ie no gauge potential) anyons described by multi valued
wave functions \cite{wu1,wilc6}.   The configuration space  is the set of
unordered $N$-tuples of complex numbers $x=\{z_1..z_N\}$ (where
$z_k=x_k+iy_k$)     of which no two are identical. In this notation the
transformation reads,
\be
\Psi & \rightarrow  & \Psi_{\theta} =
\left(\frac{\gamma}{|\gamma|}\right)^{\nu} \Psi
     \equiv g \Psi \label{eq:anyontransform} \\
A & \rightarrow & A+g^{-1}dg \nonumber
\ee
where $\nu = \theta/\pi$ and
\be
\gamma  =  \prod_{j<k}(z_j-z_k)
\label{eq:gamma}
\ee
leading to multi valued wave
functions of the form \be
\Psi (z_1..\overline{z}_N) & = & \gamma^{\nu}\Psi_B (z_1..\overline{z}_N)
\ee
where $\Psi_B$ is single valued (bosonic).

We now show that a similar construction is possible also on the
cylinder.\footnote{
A mean field Chern-Simons
description of anyons on a cylinder has been studied in ref. \cite{chak1}. }
On a cylinder of radius $R$, the points   $z_k$ and $z_k+2\pi iR$
are identified.
The configuration space $Q^N$ in (\ref{eq:QN}) is again  unordered
$N$-tuples of complex numbers but
with the identification of $z_k$ with $z_k+2\pi iR$.
The gauge transformation corresponding to (\ref{eq:anyontransform})  reads,
\be
\Psi & \longrightarrow & \Psi_{\theta,\phi} =
(\frac{\gamma_c}{|\gamma_c|})^{\nu_1}
       \exp (\frac{i\nu_2}{2R}\sum_i  )\Psi \ \ ,
\label{eq:cyltransform}
\ee
where $\nu_1 =\theta/\pi$, $\nu_2 = [\phi-(N-1)]/\pi$ and
\be
\gamma_c = \prod_{j<k}\sinh (\frac{z_j-z_k}{2R}) \ \ \ \ \  .
\label{eq:gammac}
\ee

The factor $\gamma_c$, which generalizes $\gamma$ to the cylinder, has
correct periodicity to ensure that the gauge transformation
(\ref{eq:cyltransform}) is well defined on the cylinder. For  relative
distances much smaller than $R$,   $\gamma_c$ approaches $\gamma$  and the
dependence on the phase $\phi$ disappears. Thus we recover the result
(\ref{eq:anyontransform}) appropriate to the infinite plane.

To check that (\ref{eq:cyltransform}) indeed is correct we calculate the
phases representing  the generators $\sigma_i$ and $\rho_i$ using the
multi-valued wave functions (\ref{eq:cyltransform})\footnote{If we use a
description where the gauge potential is zero the phases are encoded in the
multi-valued wave functions;  when particle coordinates are transformed
along a closed path $\Psi$ is to pick up the corresponding phase
\cite{wilc6}.} .  First we exchange particle $j$ with particle $j+1$ counter
clockwise by writing $z_j-z_{j+1}\equiv z=|z|e^{i\phi}$  and increasing  the
argument $\phi$ with $\pi$. Under this transformation  $\gamma_c\rightarrow
e^{i\pi \nu_1}\gamma_c$ while the center of mass piece in the exponential
and the bosonic part  $|\gamma_c|^{-\nu_1}\Psi_B$ are invariant so that
\be
\Psi_{\nu_1,\nu_2}  {{{\sigma_j}\atop \longrightarrow}\atop { \ }}
                        e^{i\pi \nu_1}\Psi_{\nu_1,\nu_2} \ \
\label{eq:sigmatransf}
\ee
as required.  To get $\rho_1$ we consider $z_j \rightarrow z_j+2\pi iR$ with
$Re(z_j) < Re(z_k)$ all $k\neq j$ (take particle $j$ once around the cylinder
with all other particles to the right of the path). Then
\be
\gamma_c^{\nu_1} & {{{\rho_1}\atop \longrightarrow}\atop { \ }}
 & e^{i\pi (N-1)\nu_1}\gamma_c^{\nu_1} \nonumber \\
   e^{\frac{\nu_2}{4R}\sum_i (z_i - \overline{z}_i)}
 & \ {{{\rho_1}\atop \longrightarrow}\atop { \ }} \
 & e^{\pi i\nu_2}e^{\frac{\nu_2}{4R}\sum_i(z_i-\overline{z}_i)}
\label{eq:rhochange}
\ee
so that
\be
\Psi_{\nu_1,\nu_2}  {{{\rho_1}\atop \longrightarrow}\atop { \ }}
  e^{i\pi [\nu_2 +(N-1) \nu_1]}\Psi_{\nu_1,\nu_2}
\label{eq:rhotransf}
\ee
which completes the demonstration.

\section{Cylinder with constant $\vec B$ field }
\subsection{General considerations}
Let us now put a homogeneous magnetic field $B$ perpendicular to the surface
of the cylinder. The vector potential
\be
\vec{A}(x,y) & = & (x-c)B\hat{y}
\label{eq:gaugechoice}
\ee
where $c$ is an arbitrary constant,
generates the required magnetic field and obeys the periodicity condition
$\vec{A}(x,y)=\vec{A}(x,y+2\pi R)$.
One might think that $c$ has no physical
significance, since it can be shifted by a gauge transformation
$\vec{A} \rightarrow \vec{A} + \vec\nabla \Lambda$ with
$\Lambda = y\cdot {\rm const}.$. This conclusion is incorrect however, since
$\Lambda$ is not single valued on the cylinder. Calculating the
Wilson loop $W[\Gamma ]$ for (\ref{eq:gaugechoice})
around the cylinder for fixed $x$ (Fig. 3a)
\be
\exp (iq\oint_{\Gamma_x} d\vec{r}\cdot \vec{A}) & = & e^{2\pi i qRB(x-c)}
\label{eq:Wilsonphase}
\ee
we see that a change in $c$ is equivalent to a change in the value of the
Wilson loop. Indeed, defining $e^{i\phi}$ as the Wilson loop around
$\Gamma_{x=0}$ we have
\be
\phi =  -2\pi RqBc  \ \ .
\label{eq:fluxcrel}
\ee
To understand (\ref{eq:fluxcrel}) it
is again instructive to consider a real
magnetic field configuration in three dimensions as in Fig. 3b.
Then the integral
in (\ref{eq:Wilsonphase})
measures the flux $\Phi (x)$
flowing through the cylinder
at $x$. From the figure it is clear that  $\Phi (x)$ is
linear in $x$ since  a flux of
$2\pi RB$ leaves the cylinder per unit length by crossing
the cylinder surface and
thus $c$ is the position where the net flux through the
cylinder is zero. The relation (\ref{eq:Wilsonphase})
implies that translation invariance
is broken by the Aharonov-Bohm effect, since the phase-factor is a physical
observable. It is also clear that {\it the translational symmetry is
broken down to the discrete symmetry}
$x\rightarrow x+na$, which is  also manifest in the wave functions.
Note, that adding an integer number $n$ of flux quanta through the cylinder
(\ie $\Phi (x) \rightarrow
\Phi (x) + \frac{2\pi}{q}n$) leaves the Wilson loop (\ref{eq:Wilsonphase})
invariant. Thus this does not change the physics on the surface but  shifts
$c\rightarrow c+na$.

These considerations emphasizes the importance of  boundary conditions.
The cylinder can be viewed as an infinite strip where $0\le y < 2\pi R$,
with periodic boundary conditions in $y$. If we instead had imposed, say,
hard wall boundary conditions (or confined $y$ close to a constant by some
potential), translation invariance would have remained unbroken.
It is also
necessary that the cylinder surface is penetrated by a net flux. To see this,
consider a magnetic field which is independent of $x$ but with an arbitrary
$y$-dependence.
Then if (and only if)
\be
\int_0^{2\pi R}dy B(y) =0
\ee
we can pick the gauge
\be
A_x = \int_0^y dy' B(y')
\ee
which gives a fully translationally invariant  Hamiltonian and
wave
functions;
\be
& H  =  \frac{1}{2m}\left[ (i\partial_x +
qA_x(y))^2-\partial_y^2\right] & \\
&\psi (x,y) = e^{ikx}f(y) & \nonumber
\ee
with $k$ continuous.

\subsection {One-particle solutions}
Choosing $c=0$ in (\ref{eq:gaugechoice}) the Hamiltonian is
\be
{\cal H}_0 & = & -\frac{1}{2m}\sum_j \left[ \vec{\nabla}_j
                                  -iq\vec{A}_j \right]^2 \ \ ,
\label{eq:FullHamiltonian} \\
\vec{A}_j & = & Bx_j\hat{y} \nonumber
\ee
where $j$ labels the particles. Measuring energy in units of the
cyclotron frequency $\omega_c =
\frac{qB}{m}$, the one particle eigenstates of
(\ref{eq:FullHamiltonian}) are
\be
\psi_{m,n}(x,y) & = & N_m e^{-\frac{1}{2\ell^2}(x-k_n\ell^2)^2}
e^{ik_ny}H_m(x-k_n\ell^2) \ \ , \label{eq:Fullonepsolution} \\
m=0,1,2.. & ; & n=0,\pm 1, \pm 2.. \ \ , \nonumber \\
k_n & = &  n/R \ \ , \nonumber
\ee
where $H_m$ are  Hermite polynomials, the integers $m$ labels the
Landau levels with  energies $m+\frac{1}{2}$ and the $N_m$'s are
normalization constants. The quantization of the wave number $k$ is from
demanding single valued wave functions,  $\Psi (x,y) = \Psi (x,y+2\pi R)$.
The properly normalized lowest Landau
level ($m=0$) states
\be
\psi_{n}(x,y) & = & \left( \frac{1}{4R^2 \pi^3l^2} \right)^\frac{1}{4}
                   e^{i\frac{n}{R}y}e^{-\frac{1}{\ell^2}(x-na)^2} \ \ .
\label{eq:LowestLandaustates}
\ee
are gaussians centered at positions $x_n =na$.
Notice that one cannot form any translationally invariant linear combination
of these states. Indeed, any  translated state $\psi (x-\alpha ,y)$ will be
a superposition of states from  different Landau levels and hence no longer
an eigenstate to  (\ref{eq:FullHamiltonian}).
Instead (\ref{eq:LowestLandaustates}) reflects the
discrete symmetry $x\rightarrow x+a$ discussed
above.

\section{Dimensional reduction}
\subsection{General discussion}
As discussed in the introduction, the presence of the two length scales $R$
and $\ell $ allow several possibilities for dimensional reduction, and we
will start by recalling some facts about dimensional reduction in anyon
systems. For a more comprehensive treatment, se \cite{hans6}.  Naturally we
would expect that the one dimensional model would ''remember'' the
statistics of the original particles. The situation is, however, a bit
complicated since there is no unique definition of statistics in 1+1
dimensions. One possibility is to observe that for hard core particles, we
can choose a particular ordering of the particles thus reducing the
configuration space of N particles to $R\times  R_+^{N-1}$ where $R_+$ is
the the real half line. Specifying the statistics now corresponds to
choosing a boundary condition that guarantees unitarity at the coincidence
points. As usual it is sufficient to study the two body problem, and in the
(relative) coordinate $x$ unitarity is ensured by the class of boundary
conditions \be \Psi(0) = \eta \Psi^\prime(0)  \label{eq:boundarycond} \ee
which makes the probability current vanish at $x=0$. With this definition
$\eta =0$ corresponds to bosons and $\eta^{-1}$ to fermions. This approach
has been pursued in \cite{hans6} where it is shown that when $2d$ anyons are
confined by some strong potential to move only along a line, there is a
non-trivial connection between the statistical angle $\theta$ of the anyons
and the statistical parameter $\eta$ of the resulting effective $1d$ system.
This also holds true for imposing periodic boundary conditions at $y=0$ and
$y=2\pi R$, which is of course nothing but the $R \ll \ell$ limit of the
cylinder problem studied in this paper. A calculation very similar to that
in \cite{hans6} gives, for small $R$,   \be \eta^{-1} =
R\cot^2\left(\frac{\theta}{2}\right)\frac{7\zeta (3)}{\pi^2} \sim
0.823R\cot^2\left(\frac{\theta}{2}\right)  \label{eq:etathetarel} \ee to
first order in $\cot\left(\frac{\theta}{2}\right)$ ($\zeta (z)$ being  the
Riemann  zeta-function). The flux $\Phi$ through the cylinder does not enter
(\ref{eq:etathetarel}) since for particles with identical charge, only the
center of mass movement is affected by $\Phi$; no flux flows in the
'relative' cylinder. Would we instead consider particles with different
charges (\ie an anyon anti-anyon pair) the relation (\ref{eq:etathetarel})
will have a non-trivial $\Phi$ dependence. \\

In the presence of a strong magnetic field there is another possible
dimensional reduction corresponding to the restriction to the lowest Landau
level. In the limit $\ell \ll R$ the effects of the finite radius of the
cylinder can be neglected and the general analysis in \cite{hans6} is
applicable also to the cylinder case.

However, as mentioned in the introduction, there  is non-trivial combination
of the two dimensional
reduction discussed above, namely
\be
qB &\rightarrow& \infty    \nonumber \\
R &\rightarrow& 0 \label{eq:limit} \\
\frac {\ell^2} R  &\rightarrow& a \nonumber
\ee
where $a$ will be identified as the lattice constant in the
resulting discrete model. First we study the  limit (\ref{eq:limit}) for
interacting fermions and then for non-interacting anyons.

\subsection{Fermions on the cylinder}
If the Landau energy $\hbar \omega_c$ is much larger than
the typical interaction energy between the fermions the mixing of higher
Landau levels in the lowest energy states will be suppressed. We thus
restrict the Hilbert space  to the lowest Landau level
\footnote{We assume that the number of particles is not
sufficient to fill one Landau level, \ie the filling factor is
less than one.} states .
The fermionic operators $c_n^\dagger$ and $c_n$ obey the canonical
anticommutation relations \be [ c_i ,c_j^\dagger ]_+ & = & \delta_{ij}
\label{eq:anticom}
\ee
and create and annihilate a fermion in the
$\psi_n$ state respectively. Adding a generic two body interaction
$\sum_{i\neq j} V(\vec{r}_i-\vec{r}_j)$ to (\ref{eq:FullHamiltonian}) the
full second quantized  Hamiltonian becomes
\footnote{The fermions are taken to be spinless. If
spin is included the $c_n$ operators come with an additional spin index and
a term corresponding to a Zeeman term in the Hamiltonian.}
\be
\hat{H} & = & \frac{1}{2}\sum_n c^\dagger_n c_n +
              \frac{1}{2}\sum_{ij,kl}V_{ij,kl}
              c^\dagger_ic^\dagger_jc_lc_k \ \ ; \label{eq:interham1} \\
V_{ij,kl} & = & \int d\vec{r}_1 d\vec{r}_2\,
                 \psi^*_i(\vec{r}_1)\psi^*_j(\vec{r}_2)
   V(\vec{r}_1-\vec{r}_2)\psi_k(\vec{r}_1)\psi_l(\vec{r}_2) \ \ . \nonumber
\ee
Fourier expanding in the $y$-direction
\be
V(\vec{r}) & = & \sum_{m= \ -\infty}^\infty V^m(x)e^{imy/R}
\label{eq:fouriersum}
\ee
the Hamiltonian (\ref{eq:interham1}) becomes
\be
\hat{H} & = & \half \sum_n c_n^\dagger c_n + \half \sum_{i,j,m}
              e^{-m^2\frac{a}{2R}}V^m_{i-j}c_i^\dagger c_j^\dagger
              c_{j+m}c_{i-m}
\label{eq:interham2}
\ee
where
\be
V^m_s & = & \frac{1}{\sqrt{8\pi}}\int_{-\infty}^{\infty}dx
            e^{-\frac{1}{2aR} [x-(s-m)a]^2}V^m (x) \  \ .
\label{eq:Vka}
\ee

The first term in (\ref{eq:interham2}) counts the total number of fermions
giving the magnetic zero point energy $N/2$ which acts like a chemical
potential.  The second term induces pair-wise hopping; the fermion
at site $i$ moves $m$ positions to the right while the fermion at site $j$
moves $m$ positions to the left. The hopping amplitude $V^m_{i-j}$ depends
on the distance $i-j$ between the fermions and the hopping distance $m$, see
Fig.4. The exact form of $V^m_s$ depends on the potential $V(\vec{r})$
(\ref{eq:Vka}) but the general form of the interaction part of the
Hamiltonian is easy to appreciate; since  in (\ref{eq:Fullonepsolution}) the
lattice position $n$ of a particle is related to the
canonical momentum $p_y$ through $p_y = k_n = n/R$
any single particle hopping is forbidden by conservation of total $P_y=\sum
p_y$ and hence the linear momentum along the cylinder is quenched by the
magnetic field. For the same reason any two particle interaction must move
the two particles the same number of lattice sites but in opposite
directions. From  the translational
invariance of $V(\vec{r}_1 - \vec{r}_2)$ it finally follows that the
amplitude depends only on the hopping distance $m$ and the particle
separation $i-j$. The conservation of  $P_y$ hence implies that
there can be no net charge transport along the cylinder.

In the limit (\ref{eq:limit}), all $m\neq 0$ contributions get exponentially
damped and we are left with,
\be
\hat{H} & = & \frac{1}{2}\sum_i n_i + \half\sum_{i\neq j}V^0_{i-j}n_i n_j
\ \ \ \ \ ,
\label{eq:interham3}
\ee
so any distinct distribution of fermions on the lattice sites is an
eigenstate. The system is classical in the sense that
the kinetic energy only contributes the constant $N/2$ (which vanishes
when $\hbar\rightarrow 0$) so that the Hamiltonian is purely potential
and lacks dynamics. From general arguments follows that at
finite temperature all lattice sites are
occupied with equal probability and the correlation function
$C_{ij} =  \hat{K}_{i-j} = \langle n_i n_j\rangle - \langle n_i
\rangle\langle n_j  \rangle$ decays
exponentially, \ie there is no spontaneous symmetry breaking. This will
however {\it not} destroy the lattice structure itself since the lattice is
not formed dynamically by inter-particle forces but instead enforced by the
external gauge potential. Indeed, shifting any combination of the
eigenstates (\ref{eq:LowestLandaustates}) by a fraction of the lattice
constant $a$ will involve the mixing of higher higher Landau level states
and thus increase the energy. The periodicity of the lattice hence persists
at finite temperature and may be detected as a periodic charge density.

The Hamiltonian (\ref{eq:interham3}) is of course nothing but the familiar
lattice gas
model and if only nearest neighbour interaction is included
we recover the $1d$ Ising chain through the identification $S^z_i = n_i
-\half$ \cite{huan1}.

\subsection{Anyons on the cylinder}
It is easily shown that {\it any} function
\be
\Psi (x_j,y_j) = f(z_1..z_N)\exp\left[-\frac{1}{2\ell^2}\sum_j x_j^2\right]
\ee
where $f(z_j)$ is holomorphic in $z_j=x_j+iy_j$ is an eigenfunction to
the Hamiltonian (\ref{eq:FullHamiltonian}) with energy $N\hbar\omega_c/2$.
{}From this and (\ref{eq:cyltransform}) we can immediately find the anyonic
wave functions corresponding to the lowest Landau level on a cylinder:
\footnote{
With anyons in the lowest Landau level we mean anyonic states continuing
Bose states where all particles are in the lowest Landau level \cite{karl1}. }
\be
\Psi_{\{n_j\} }(x_j,y_j) = \gamma_c^\nu S\left\{ \exp \sum_{j=1}^N \left[
i\frac{n_j}{R}y_j + \frac{1}{2Ra}(x_j - an_j)^2 \right] \right\}
\label{eq:anyonwf}
\ee
Here $\gamma_c$ is given by (\ref{eq:gammac}) and $S$ means symmetrization
of the particle indices. It is interesting to ask if there
is some effective one dimensional model corresponding to these solutions.
If the anyons move in an infinite plane this is indeed the case and the
equivalent model is the Calogero model, \ie particles interacting via
pairwise harmonic and $1/x^2$ forces\cite{calo1,calo2}. The Calogero model
is integrable,  and the spectrum (energies\cite{polyc1} and wave
functions\cite{brin3}) of the  Calogero Hamiltonian is
the same as that for the conserved total angular momentum operator for the
$N$ anyon system
\footnote{Note that it is the conserved total {\it angular momentum}
operator in the  lowest Landau level anyon system which corresponds to the
Hamiltonian of the Calogero model. The spectrum of the {\it Hamiltonian} in
the former system is of course trivial since all lowest Landau level states
have the same energy of $N\hbar\omega_c/2$.}.

In the case of a cylinder there is in general no such correspondence,
but we shall show that in the limit (\ref{eq:limit})
the anyon problem is equivalent to another integrable one-dimensional
quantum system, namely the Sutherland model on a circle. The Sutherland
model is defined by the Hamiltonian
\be
H_{Suth} = -\sum_{j=1}^N \partial^2_{\vartheta_j} + \sum_{j\neq k}
      \frac g   {\sin^2(\vartheta_j - \vartheta_j)}
\label{eq:Hsuth}
\ee
where the angular variables $\vartheta_j$  range from 0 to $2\pi$
\cite{suth3,suth2}. The spectrum of (\ref{eq:Hsuth}) is given by
\be
E_{\{n_j\} } = 4\left\{ \sum_j n_j^2
+\lambda\sum_{j>k}(n_j-n_k)+\frac{\lambda^2}{12}
               N(N^2-1)\right\}
\label{eq:suthspec}
\ee
where $\lambda = \half [1+\sqrt{1+2g}]$ and the quantum numbers $n_j$ are
ordered integers with $n_j > n_k$ for $j>k$ \cite{suth3}.
To understand the connection to anyons on a cylinder it is useful to
consider the guiding center coordinates for the cylinder problem, defined as
\be
X_j &= & x_j + \frac m {qB} v_{y_j} = \ell^2 p_{y_j}  \label{eq:guiding} \\
Y_j &= & y_j - \frac m {qB} v_{x_j} = y_j - \ell^2 p_{x_j}
                                \nonumber
\ee
with the commutator
\be
[X_j, Y_k] = -i\ell^2 \delta_{jk}
\ee
In the plane, using radial gauge, the conserved (canonical) angular momentum
$L$, can be
expressed as $L=\frac 1 {2\ell^2} \sum_i(X_i^2 + Y_i^2) - H/\omega_c$, and
in the lowest Landau level, the last term is just $1/2$. This amounts to
observing that,  in the radial
gauge, the guiding center radius $R_i^2 = X_i^2 + Y_i^2$ is essentially the
angular momentum, and thus a constant of motion. It is this conserved
quantity that,  can be
reinterpreted as the Hamiltonian for the equivalent one dimensional model -
in this case the Calogero model. On the cylinder, the (canonical) angular
momentum is not conserved, but we  have another conserved quantity, namely
the total momentum in the $y$-direction, which according to
(\ref{eq:guiding}) is nothing but the $x$-component of the guiding center
coordinate. Note that since $Y_j \in [0, 2\pi]$,  $X_j$ is quantized,
\be
X_i^{n_i} = n_i \frac {\ell^2} R \ \ .
\ee
Now we have a very natural guess for the 'Hamiltonian' for the equivalent
one dimensional model, namely
\be
H_e = \frac {1} {2\ell^2} \sum_{j=0}^N X_j^2 =
-\frac {\ell^2} 2 \sum_{j=0}^N  \frac {\partial^2} {\partial Y_j^2} =
-\frac {\ell^2} 2 \sum_{j=0}^N  \frac {\partial^2} {\partial y_j^2}
                     \ \ .
\label{eq:limitham}
\ee
$H_e$ commutes with the Hamiltonian and it is the    $Y_i \rightarrow 0$
limit of the quantity $L=\frac 1 {2\ell^2} \sum_i(X_i^2 + Y_i^2)$ discussed
above.  The eigenfunctions of $H_e$ in the $Y$ representation (where $X_j =
(i\ell^2)  \frac{\partial_j}{\partial Y_j}$ are given by
\be
\Psi_{\{n_j\} }(Y_j) = N \exp \sum_{j=1}^N \left[ i\frac {n_j} R Y_j \right] =
       N \exp \sum_{j=1}^N \left[ i\frac {n_j} R y_j -i n_j
       \frac{\ell^2}{R} p_{x_j} \right]
\ee
In the $(x,y)$ representation this becomes
\be
\Psi_{\{n_j\} }(x_j, y_j) =  N \exp\left( \sum_{j=1}^N \left[ i\frac {n_j}
R y_j\right]\right) \prod_{j=1}^N \delta(x_j - n_j a)
\ee
Note that these wave functions can also be obtained from
(\ref{eq:Fullonepsolution})
by taking the limit (\ref{eq:limit}), and using the representation of the
delta function as a limit of a gaussian.

We now turn to the anyon problem. If we order the $x$ coordinates so
$x_j<x_k$ for $j<k$, it is easy to show that up to an unimportant overall
phase factor
\be
 \gamma_c \rightarrow \prod_{j>k} e^{\frac {x_j-x_k} {2R} }
       e^{i\nu \frac {y_j -y_k} {2R} }\ \ \ \ \ {\rm for}\  R\rightarrow 0
\ee
so the anyonic wave function (\ref{eq:anyonwf}) becomes
\be
\Psi_{\{n_j\} }(x_j, y_j) =  N \exp\left( \sum_{j=1}^N \left[ i\frac {n_j}
R y_j \right] + i\nu \sum_{j<k} \frac {y_j -y_k} {2R} \right)
          \prod_{j=1}^N \delta(x_j - n_j a)
\label{eq:limitwf}
\ee
After some combinatorics, it follows that (\ref{eq:limitwf}) are
indeed eigenfunctions for the effective Hamiltonian $H_e$ in
(\ref{eq:limitham}) and that the spectrum  is
\be
E_{\{n_j\} } = \frac {\ell^2} {2R^2}\left[ \sum_{j=1}^N n_j^2 +
\nu\sum_{j>k}(n_j-n_k) + \nu^2 \frac {N(N^2-1)}{12} \right]
\ee
which is exactly the spectrum (\ref{eq:suthspec}) obtained by Sutherland,
if we identify $\lambda = \nu$ and measure ''energy'' in units of $\frac
{\ell^2} {2R^2}$. Note that it follows from the delta function in
(\ref{eq:limitwf}) and the assumption of the ordering of the $x_j$ that the
ordering of the quantum numbers $n_j$ is the same as in the Sutherland
spectrum (\ref{eq:suthspec}).

When it came to energy eigenvalues and degeneracies, it was fairly easy to
establish the  equivalence between the Calogero model and anyons in the
lowest Landau level, both for the infinite plane and the cylinder. It is
much harder to obtain the corresponding connection between the wave
functions.  In fact, this has been done only recently \cite{brin3} in the
case of the plane, using an operator approach which relies heavily on  that
the spectrum  can be generated by step  operators.  Since the spectrum of
the Sutherland model is not equally spaced we cannot expect any simple
generalization of this method to be applicable, and we have not been able to
understand the connection between the wave functions.  Nevertheless, the
equivalence of the spectra strongly indicates that in the  limit
(\ref{eq:limit}), anyons on a cylinder is just another representation of the
Sutherland model, and with the advantage of having very simple wave
functions.  Finally we note that the proof of the equivalence between the
anyon system  and the Calogero model given in \cite{brin3} is based on the
observation  that they are just two equivalent representations of the same
algebra. Thus  the proof of equivalence did not require the knowledge of the
wave functions.  In \cite{polyc4} Polychronakos has given the  algebra
underlying the Sutherland model, and it would be interesting to check
whether the same algebra is realized by the anyon problem considered above.
It is also interesting to speculate what happens if we compactify both
directions, \ie consider anyons on a torus. In this case there are only a
finite number of states in the lowest Landau level, so we expect the
corresponding equivalent one-dimensional model to be a lattice model with a
finite number of points determined by the total flux out of the cylinder. In
ref. \cite{hans6} we guessed that such a model would be of the type
considered by Haldane \cite{hald2,hald3} and Shastry \cite{shas1}, who
constructed a lattice version of the Sutherland model. The results of this
paper supports that guess, and we can furthermore assert that the
correspondence between anyons on a torus and a one dimensional lattice model
should be looked for in the limit (\ref{eq:limit}).

\section{Order of magnitude estimates and experimental implications}
In this section we ask whether any of the effects above could be detected
experimentally. Having in mind the  Buckytubes \cite{iiji1} we may also
ask whether a microscopic cylindrical topology
could affect any magnetic properties of a macroscopic sample put
in an external magnetic field.
Recall from the end of section 3.1 that in order to get flux
induced breakdown
of the translational symmetry there must be a net magnetic flux passing
through the  cylinder surface. It should be obvious
that a homogeneous magnetic field will not do and that what is needed is
rather something like the field at the end of a thin solenoid. This almost
immediately  rules out
the possibility to see any effect in a system of Buckytubes since this
would  require a magnetic field which is strongly inhomogeneous on the
length scale of  $1$~nm.
\footnote{In fact, even to get a quantum of magnetic flux into the buckytube
would require a magnetic field $\sim 100T$.}
Ignoring for a moment the technical
difficulties of actually taking the magnetic field out through the
surface of the cylinder we now estimate on which scales a
lattice can form. \\

For the lattice to be significant we must require that $\ell < a$, \ie
$R < \ell $ or $qBR^2 < 1$.
For given radius $R$ this puts an upper bound
on the field strength $B_\perp $. For the lattice structure to be of {\it
dynamical} significance the  magnetic energy must in addition
be larger than (or at least of the same
magnitude as) the typical (Coulomb) interaction energy between the particles
and for the Landau level structure to be robust against thermal
fluctuations $kT \leq \hbar\omega_c$
which puts an absolute lower
bound on $B_\perp $.
\footnote{If the electron system is
to be described by a 1d lattice model this condition sharpens to $kT <<
\hbar\omega_c$ so that the lowest Landau level approximation is valid.
Nevertheless the weaker condition $kT \leq \hbar\omega_c$ is sufficient for
the Landau level structure to survive and a charge density to form.} Finally
one magnetic flux quantum per lattice site must be inserted at the end of
the cylinder in order to be taken out homogeneously along the cylinder.
Ignoring the difficulties of extracting the magnetic flux  smoothly through
the cylinder surface these conditions summarize to \be
\pi R^2 B_\parallel & \geq & N\Phi_0 \label{eq:Condition1} \\
R & \leq & a \label{eq:Condition2} \\
kT & \leq & \hbar\omega_c \label{eq:Condition3} \\
n & \leq & \rho_{LL} \label{eq:Condition4}  \ \ \ \ \ .
\ee
Here $B_\parallel$ is the field strength
injecting a flux $\Phi =\pi R^2 B_\parallel$ into the cylinder, $N$ is the
number of lattice sites and $n$ the charge carrier density.
Of course, we also assume that the gap to transverse excitations is big
enough  for the system to be effectively two-dimensional.
Starting with the lower bound on the magnetic field and comparing with the
QHE the Landau level structure is certainly of importance at magnetic field
strengths around $0.01-1$T and temperatures below $1$K.
Taking $R\sim a_0$, the largest possible value of $R$, creating a
lattice of the order of a couple of hundred lattice sites would require
\be
B_\parallel & \sim & 100B_\perp  \ \ .
\label{eq:ConditionBMAX}
\ee
Thus $B_\parallel\sim 10$T  when radius $R\sim a\sim 10^{-7}$m. An even
smaller radius, permitted by (\ref{eq:Condition2}) would increase
$B_\parallel$ through (\ref{eq:Condition1}). This effectively rules out the
possibility for a lattice structure to form on buckytubes (where the typical
radii are of the order of one nm).
While the
radius $R\sim a_0\sim 10^{-7}$m, implying a layer thickness $\Delta z \sim
10$\AA,  is not unrealistic
the difficulty in
taking out the injected flux through the surface of the cylinder
makes an experiment presently impossible, since already the
required $B_\parallel$ is on the limit of accessible magnetic fields.
\vskip 2mm\noi
{\bf Acknowledgement:} We thank Anders Karlhede for many interesting
discussions, and for a critical reading of the manuscript.

\newpage
{\center \bf Figure captions}

{\bf Fig 1a:} The generators to the braid group on the cylinder; $\sigma_i$

exchanges particles $i$ and $i+1$ counter clockwise and $\rho_k$ takes
one particle once around the cylinder, leaving $k-1$ particles to the left
of the path. \\

{\bf Fig 1b:} The trajectory corresponding to the group
element $\sigma_i\rho_i\sigma_i$ can continuously be transformed into the
trajectory representing $\rho_{i+1}$, thus implying the relation
(\ref{eq:generatorref4}). \\

{\bf Fig 2:} If the phases associated with the braid group generators are
represented by Wilson loops, the phase $e^{i\theta}$ is induced by a
statistical
magnetic flux of $2\theta$ tied to each particle. From the picture it is
clear that the
flux enclosed by the path $\rho_k$ depends on $k$ as in (\ref{eq:rhokrel}).
\\

{\bf Fig 3a:} The path $\Gamma_x$. \\

{\bf Fig 3b:} A real magnetic field configuration generating a constant
magnetic
flux through
the surface of the cylinder. The flux, as measured by the Wilson loop
$\Gamma_x$, increases linearly
in $x$ and $x=c$ is the point where this is zero. \\

{\bf Fig 4:} The two particle hopping induced by $V^m_{i-j}$.

\end{document}